# CaraNet: Context Axial Reverse Attention Network for Segmentation of Small Medical Objects

Ange Lou[1], Shuyue Guan[2], Hanseok Ko[3], Murray Loew[2]
[1]Vanderbilt University, Nashville TN, USA
ange.lou@vanderbilt.edu
[2]The George Washington University, Washington DC, USA
frankshuyueguan@gwu.edu, loew@gwu.edu
[3]Korea University, Seoul, Korea
hsko@korea.ac.kr

**Abstract**

Segmenting medical images accurately and reliably is important for disease diagnosis and treatment. It is a challenging task because of the wide variety of objects' sizes, shapes, and scanning modalities. Recently, many convolutional neural networks (CNN) have been designed for segmentation tasks and achieved great success. Few studies, however, have fully considered the sizes of objects, and thus most demonstrate poor performance for small objects segmentation. This can have a significant impact on the early detection of diseases. This paper proposes a Context Axial Reserve Attention Network (CaraNet) to improve the segmentation performance on small objects compared with several recent state-of-the-art models. We test our CaraNet on brain tumor (BraTS 2018) and polyp (Kvasir-SEG, CVC-ColonDB, CVC-ClinicDB, CVC-300, and ETIS-LaribPolypDB) segmentation datasets. Our CaraNet achieves the top-rank mean Dice segmentation accuracy, and results show a distinct advantage of CaraNet in the segmentation of small medical objects.
Codes available: https://github.com/AngeLouCN/CaraNet

## 1. INTRODUCTION

Deep learning has had a tremendous impact on various fields in science. Our focus of the current study in deep learning is on one of the most critical areas of computer vision: medical image segmentation. Recently, various convolutional neural networks (CNNs) have shown great performance on medical image segmentation [1,2,3,4]. Those CNNs have been introduced for various medical imaging modalities, including X-ray, visible-light imaging, magnetic resonance imaging (MRI), positron emission tomography (PET), and computerized tomography (CT). They all achieved excellent performance on medical image segmentation challenges from different modalities, like BraTS [5,6,7], KiTS19 [8], and COVID19-20 [9,10]. To obtain more accurate segmentation results, many works introduced improvements of network architectures. Those improvements are mostly attributed to exploring new neural architectures by designing networks with varying depths (ResNet [11]), widths (ResNeXt [12]), connectivity (DenseNet [13] and GoogLeNet [14]), or new types of components (pyramid scene [15] and atrous convolution [16]). Although those new architectures improve the overall segmentation results, they are less sensitive to detecting small medical objects. And it is very common in medical image segmentation that the anatomy of interest occupies only a very small portion of the image [17]. Most extracted features belong to the background, while these small lesion areas are important for early detection and diagnosis. For example, the survival rate decreases with the growing size of a brain tumor [18]. Thus, it has clinical significance to build an effective network to detect tiny medical objects.

The attention mechanism plays a dominant role in neural network research. It can effectively use information transferred from several subsequent feature maps to detect the salience features [19]. Many attention methods such as self-attention and multi-head attention have been verified to have high performance in applications of natural language processing [20] and computer vision [21]. Those attention methods also have been successfully used for medical image segmentation; for example, the Medical Transformer [22] (MedT) used a gated axial self-attention layer to build a Local-Global (LoGo) network for ultrasound and microscopy image segmentation, TransUNet [23] stacked self-attention as a transformer in the encoder for CT image segmentation, and CoTr [24] bridged two CNN encoder and decoder by the transformer encoder for multi-organ segmentation. All those attention-based segmentations achieve significant improvement compared with purely convolutional neural networks like U-Net [1] and FCN [25].

Although those new types of neural networks show good performance on many medical segmentation tasks, they seldom consider the small object segmentation, especially in the medical image area. We propose here a novel attention-based deep neural network, called **C**ontext **A**xial **R**everse **A**ttention **Net**work (**CaraNet**). The contribution of the paper can be summarized as follows:
1) We propose a novel neural network – CaraNet -- to solve the problem of segmentation of small medical objects.
2) We introduce a method to evaluate the network's performance on small medical objects.
3) Our experiments show that CaraNet outperforms most current models (e.g., PraNet from MICCAI '20) and advances the state-of-the-art by a large margin, both overall and on small objects, in segmentation performance on polyps.

## 2. METHOD

Figure 1 shows the architecture of our CaraNet, which uses a parallel partial decoder [26] to generate the high-level semantic global map and a set of context and axial reverse attention operations to detect global and local feature information. We will introduce each component in the following subsections.

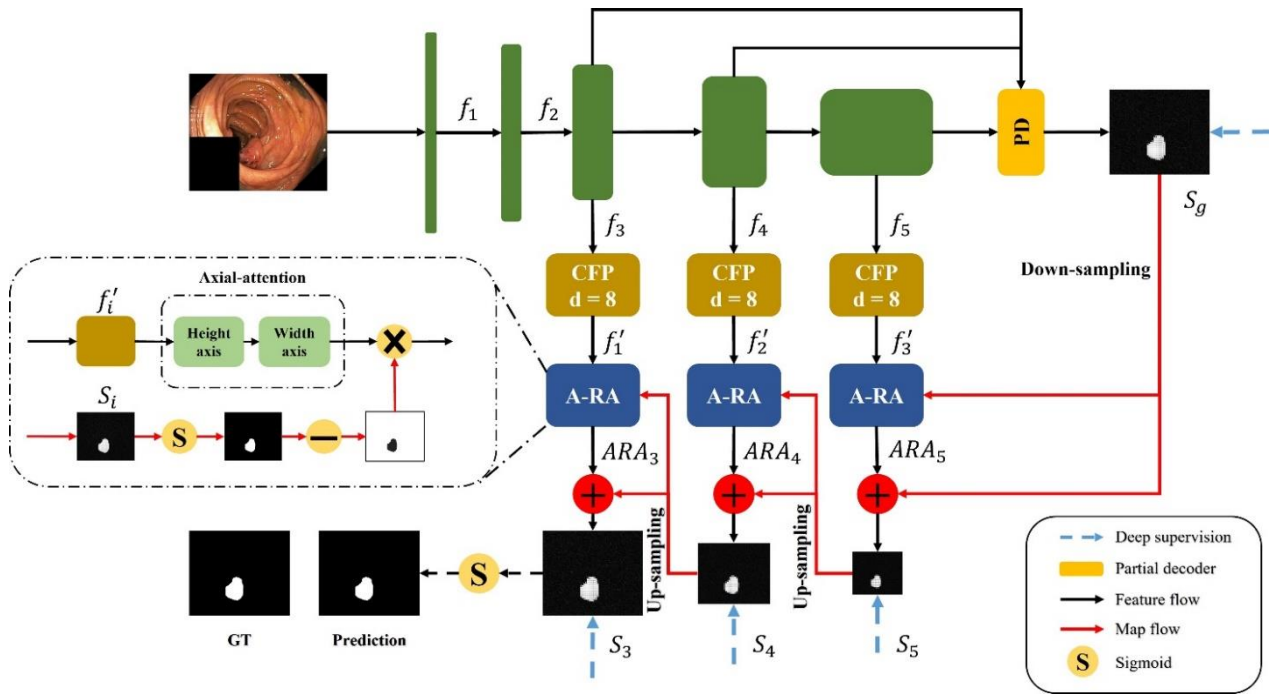

Figure 1. Overview of CaraNet, which contains pretrained backbone, partial decoder, channel-wise feature pyramid (CFP) module and axial reverse attention (A-RA) module.

### 2.1 Backbone

Transfer learning provides a feasible method for alleviating the challenge of data-hunger, and it has been widely applied to the field of computer vision [27]. Benefiting from the strong visual knowledge distributed in ImageNet [28], the pre-trained CNNs can be fine-tuned with a small amount of task-specific data and can perform well on downstream tasks. Since Res2Net [29] can construct hierarchical residual-like connections within one single residual block that has stronger multi-scale representation ability, we applied the pre-trained Res2Net as the backbone of

CaraNet.

**2.2 Partial decoder**

Existing state-of-the-art segmentation networks rely on aggregating multi-level features from the encoder (e.g., U-Net aggregates all level features extracted from an encoder). Compared to the high-level features, however, low-level features contribute less to performance but have higher computational cost because of their larger spatial resolution [30]. Thus, we applied the parallel partial decoder [26] as shown in Figure 2 to aggregate high-level features. We feed the original image which size is $h \times w \times c$ ($h$, $w$, and $c$ represent the height, width, and channel) into Res2Net, and we can get five different level features $\{f_i, i = 1, ..., 5\}$ with resolution $\{\frac{h}{2^{i-1}}, \frac{w}{2^{i-1}}\}$. We aggregated the high-level features $\{f_3, f_4, f_5\}$ from Res2Net by using the partial decoder with a parallel connection. Then, we can get a global map $S_g = PD(f_3, f_4, f_5)$.

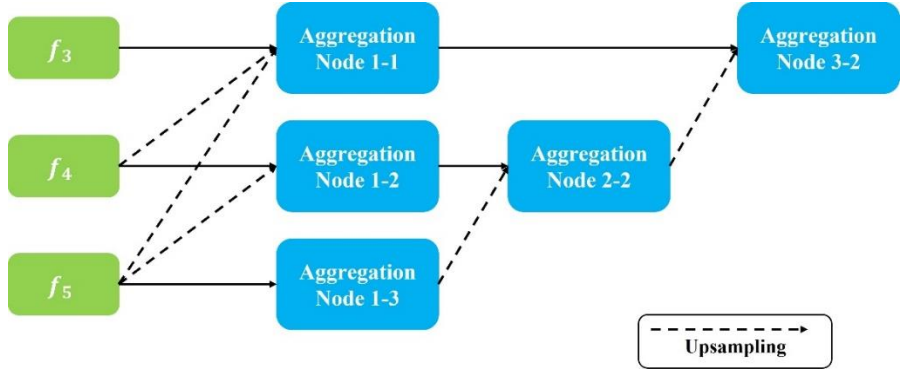

Figure 2. Overview of partial decoder with parallel connection.

**2.3 Channel-wise feature pyramid module**

The Feature Pyramid (FP) has been widely used in deep learning models for computer vision tasks due to its ability to represent multi-scale features. For example, PSPNet [31] builds a pyramid pooling module with different sizes' pooling layers to extract multi-scale features, and the Feature Pyramid Network (FPN) [32] takes different strides with convolution kernels to obtain a FP. Although those FP-based methods perform well in the computer vision area, they cannot avoid using large numbers of parameters, which consume a large amount of computation resources. Alternatively, our previous works [33, 34] proposed a lightweight Channel-wise Feature Pyramid (CFP) module and successfully applied it to both nature and medical image segmentation. The architecture of this CFP module is shown in Figure 3.

Figure 3(a) shows the architecture of the CFP module; it contains total $K$ channels and each channel has its own dilation rate $r_K$. Typically, we choose the $K = 4$ for CaraNet and the dilation rates for each channel $\{r_1, r_2, r_3, r_4\} = \{1,2,4,8\}$; thus, each channel's dimension is $M/4$. Then, we applied hierarchical feature fusion [35] (HFF) to sum the outputs of all channels step by step. For the FP channel, we provide two versions with regular convolution and asymmetric convolution as shown in Figure 3(b) and (c). We connected the outputs of each convolutional module by using skip connection, and thus each channel can be considered as a sub-pyramid. We selected the regular convolution as FP channel for CaraNet. The overall FP is obtained from concatenating those sub-pyramids from the hierarchical feature fusion operation. The final FP contains four levels of feature stacks as shown in Figure 4. These four levels of feature stacks $\{level_i, i = 1, ..., 4\}$ are computed by:

$$\begin{cases} level_1 = out_{FP1} \\ level_2 = level_1 + out_{FP2} \\ level_3 = level_2 + out_{FP3} \\ level_4 = level_3 + out_{FP4} \end{cases} \quad (1)$$

And the final FP is computed by $\sum_i level_i$.

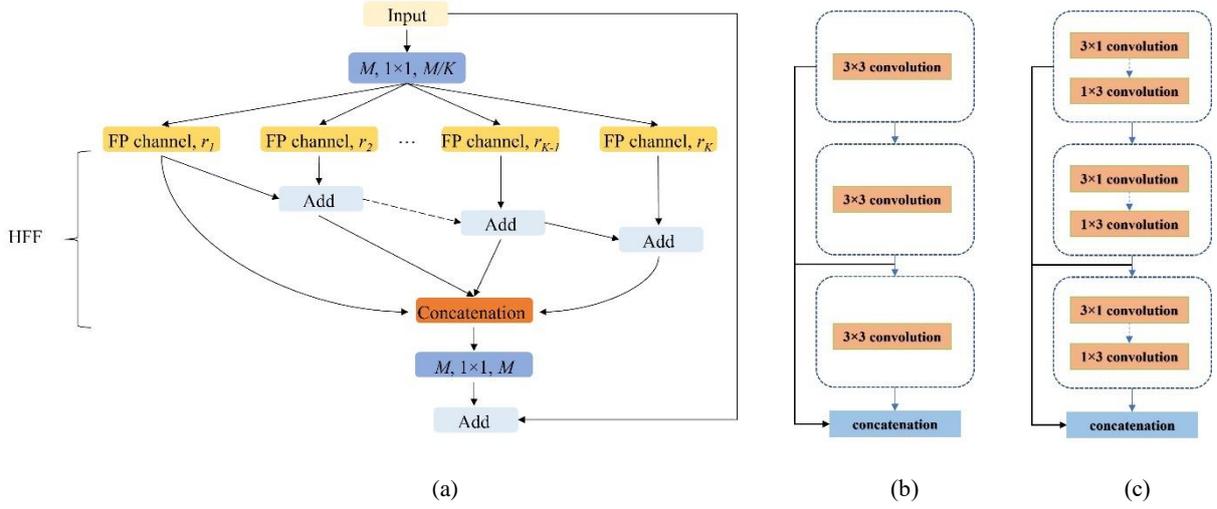

Figure 3. (a) CFP module, (b) FP channel with regular convolution, (c) FP channel with asymmetric convolution

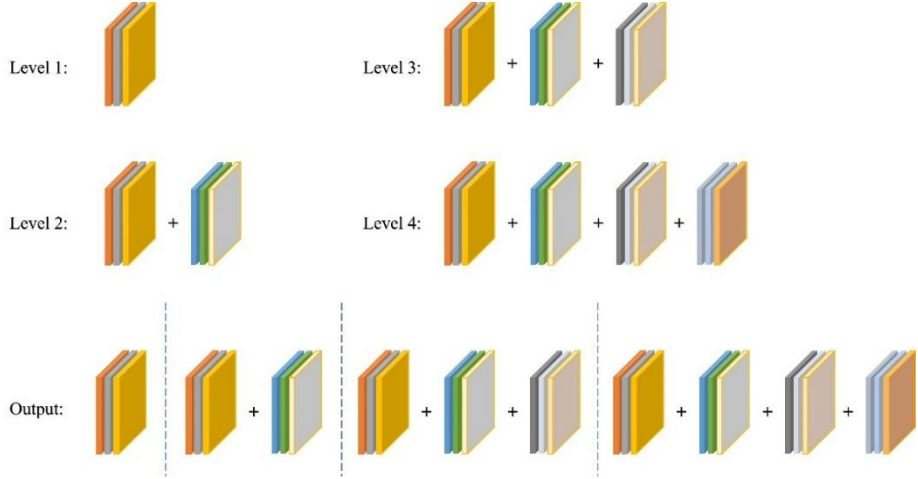

Figure 4. Final feature pyramid obtained from CFP module.

**2.4 Axial reverse attention module**

The previous partial decoder that generates the global map $S_g$ (Sec. 2.2) could roughly locate the position of medical objects, and the CFP module extracted only multi-scale features from the pre-trained model. To obtain more accurate feature information, we designed the Axial Reverse Attention (A-RA) module to analyze localization information and multi-scale features. The overview and detail of the A-RA module can be seen in Figure 1 and Figure 5, respectively. The input of the top line is the multi-scale feature maps $f_i'$ from the CFP module and we used axial attention to analyze the salience information. The axial attention is based on self-attention, which maps a query and a set of key-value pairs to an output and the operation:

$$Attention(Q, K, V) = Softmax\left(\frac{QK^T}{\sqrt{d_K}}\right) \qquad (2)$$

where $Q, K, V$, and $d_K$ represent query, key, value, and dimension of key, respectively. However, self-attention consumes great computational resources, especially when the spatial dimension of the input is large [36]. Therefore, we applied axial attention, which factorizes 2D attention into two 1D attention along height and width axes. Here we replace the softmax activation function with a sigmoid, based on the experiments. For the second line, we applied the reverse operation [37] to detect the salience features from the side-output $S_i$, which is obtained from the output of the previous A-RA module. The reverse operation is:

$$R_i = 1 - Sigmoid(S_i). \tag{3}$$

The total axial reverse attention operation is:

$$ARA_i = AA_i \odot R_i \tag{4}$$

where $\odot$ is element-wise multiplication, and the $AA_i$ is feature from the axial attention route.

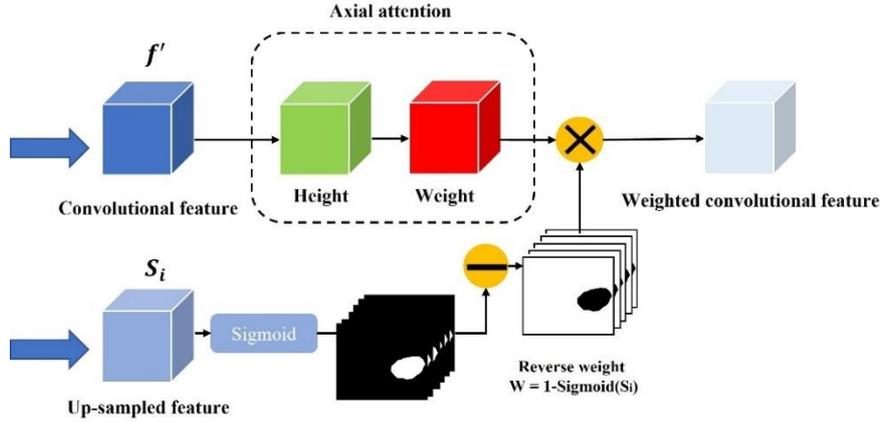

Figure 5. Structure of Axial Reverse Attention module.

## 2.5 Deep supervision

We apply weighted intersection over union (IoU) and weighted binary cross-entropy (BCE) in our loss function: $\mathcal{L} = \mathcal{L}_{IoU}^w + \mathcal{L}_{BCE}^w$ to calculate the global loss and local (pixel-level) loss, respectively. To train CaraNet, we apply deep supervision for the three side-outputs $(S_1, S_2, S_3)$ and the global map $S_g$. Before calculating the loss, we up-sampled them to the same size as ground truth $G$. Thus, the total loss:

$$\mathcal{L}_{total} = \mathcal{L}(G, S_g^{up}) + \sum_{i=3}^{5} \mathcal{L}(G, S_i^{up}) \tag{5}$$

## 2.6 Small object segmentation analysis

Since the size of all images input to the segmentation models must be fixed, the size of an object is determined by the number of pixels in the object *m* and the number of total pixels in the image *N*. Thus, we consider the object's size using the size ratio (proportion) = *m*/*N*. Then, we evaluate the performance of segmentation models according to the sizes of objects. Especially, we mainly focus on the small areas whose size ratios are smaller than 5%.

For a segmentation model, we first obtain the mean-Dice coefficients and size ratios of segmentations from the test dataset. Similar to computing the histogram, we plot the results in a curve whose y-axis is mean-Dice coefficients and x-axis is increasingly sorted size ratios. To smooth the curve, we take interval-averaged mean-Dice coefficients by sorted size ratios: we divide the entire range of size ratios into a consecutive, non-overlapping, and of equal length series of intervals, and then calculate the average mean-Dice coefficients of size ratios in each interval. The interval-averaged coefficients have a smooth curve and are more stable in the presence of noise.

## 3. EXPERIMENT

### 3.1 Implementation details

We implemented our model in PyTorch accelerated by the NVIDIA RTX GPU. We resized input images to $352 \times 352$ for polyp segmentation and $256 \times 256$ for brain tumor segmentation, and employed a multi-scale training strategy $\{0.75, 1.0, 1.25\}$ instead of data augmentation. We used Adam optimizer with the initial learning rate $1e^{-4}$.

### 3.2 Dataset

We test our CaraNet on five polyp segmentation datasets: ETIS [38], CVC-ClinicDB [39], CVC-ColonDB [40], EndoScene [41], and Kvasir [42]. The first four are standard benchmarks, and the last one is the largest dataset, which was released recently. We also test our model on a brain tumor segmentation dataset (BraTS 2018), which contains more extremely small medical objects. Table 1 shows the details of these datasets: image size, scale of testing set, and size ratios of medical objects.

Table 1. Details of datasets

|  | Image size | Number of test samples | Object size ratio |
|---|---|---|---|
| **ETIS** | $966 \times 1225$ | 196 | 0.11% - 29.05% |
| **CVC-ClinicDB** | $288 \times 384$ | 62 | 0.34 % - 45.88% |
| **CVC-ColonDB** | $500 \times 574$ | 380 | 0.30% - 63.15% |
| **CVC-300** | $500 \times 574$ | 60 | 0.55% - 18.42% |
| **Kvasir** | $1070 \times 1348$ | 100 | 0.79% - 62.13% |
| **BraTS 2018** | $256 \times 256$ | 3231 | 0.01% - 4.91% |

### 3.3 Baseline

We compared CaraNet with six medical image segmentation models, including state-of-the-art models: U-Net [1], U-Net++ [2], ResUNet-mod [43], ResUNet++ [3], SFA [44], and PraNet [27].

### 3.4 Training and measurement metrics

We randomly split 80% of images from Kvasir and CVC-ClinicDB as training set and the remainder as a testing dataset. In addition to mean Dice and mean IoU, we also apply four other measurement metrics: weighted dice metric $F_\beta^w$, MAE, enhanced alignment metric $\mathrm{E}_\phi^{max}$ [45], and structural measurement $S_\alpha$ [46]. Table 2 shows the polyp segmentation on the five datasets. The weighted dice metric $F_\beta^w$ is used to amend the "equal importance flaw" in dice. The MAE is used to measure the pixel-to-pixel accuracy. The recently released enhanced alignment metric $\mathrm{E}_\phi^{max}$ is utilized to evaluate the pixel-level and global-level similarity. And $S_\alpha$ is used to measure the structure similarity between predictions and ground truth.

### 3.5 Results

We also show some polyp segmentation results in Figure 6. For the five polyp datasets, CaraNet not only outperforms the compared models in overall performance, but also on samples with small polyps. Figure 7 shows the segmentation performance of CaraNet and PraNet for small objects (proportions $\leq 5\%$). For the extremely small object segmentation analysis on the BraTS 2018 dataset, we compare only CaraNet with PraNet because PraNet has the closest performance to ours, and the overall accuracies of the other segmentation models are clearly lower than those of CaraNet and PraNet. (Note: the fluctuations with size in colonoscopy datasets are caused by types and boundary of polyps, and quality of imaging)

Table 2. Quantitative results on Kvasir, CVC-ClinicDB, CVC-ColonDB, ETIS, and CVC-T (test dataset of EndoScene).

| | Methods | mean Dice | mean IoU | $F_\beta^w$ | $S_\alpha$ | $E_\phi^{max}$ | MAE |
|---|---|---|---|---|---|---|---|
| Kvasir | UNet | 0.818 | 0.746 | 0.794 | 0.858 | 0.893 | 0.055 |
| | UNet++ | 0.821 | 0.743 | 0.808 | 0.862 | 0.910 | 0.048 |
| | ResUNet-mod | 0.791 | n/a | n/a | n/a | n/a | n/a |
| | ResUNet++ | 0.813 | 0.793 | n/a | n/a | n/a | n/a |
| | SFA | 0.723 | 0.611 | 0.670 | 0.782 | 0.849 | 0.075 |
| | PraNet | 0.898 | 0.840 | 0.885 | 0.915 | 0.948 | 0.030 |
| | **CaraNet** | **0.918** | **0.865** | **0.909** | **0.929** | **0.968** | **0.023** |
| CVC-ClinicDB | UNet | 0.823 | 0.755 | 0.811 | 0.889 | 0.954 | 0.019 |
| | UNet++ | 0.794 | 0.729 | 0.785 | 0.873 | 0.931 | 0.022 |
| | ResUNet-mod | 0.779 | n/a | n/a | n/a | n/a | n/a |
| | ResUNet++ | 0.796 | 0.796 | n/a | n/a | n/a | n/a |
| | SFA | 0.700 | 0.607 | 0.647 | 0.793 | 0.885 | 0.042 |
| | PraNet | 0.899 | 0.849 | 0.896 | 0.936 | 0.979 | 0.009 |
| | **CaraNet** | **0.936** | **0.887** | **0.931** | **0.954** | **0.991** | **0.007** |
| ColonDB | UNet | 0.512 | 0.444 | 0.498 | 0.712 | 0.776 | 0.061 |
| | UNet++ | 0.483 | 0.410 | 0.467 | 0.691 | 0.760 | 0.064 |
| | SFA | 0.469 | 0.347 | 0.379 | 0.634 | 0.765 | 0.094 |
| | PraNet | 0.709 | 0.640 | 0.696 | 0.819 | 0.869 | 0.045 |
| | **CaraNet** | **0.773** | **0.689** | **0.729** | **0.853** | **0.902** | **0.042** |
| ETIS | UNet | 0.398 | 0.335 | 0.366 | 0.684 | 0.740 | 0.036 |
| | UNet++ | 0.401 | 0.344 | 0.390 | 0.683 | 0.776 | 0.035 |
| | SFA | 0.297 | 0.217 | 0.231 | 0.557 | 0.633 | 0.109 |
| | PraNet | 0.628 | 0.567 | 0.600 | 0.794 | 0.841 | 0.031 |
| | **CaraNet** | **0.747** | **0.672** | **0.709** | **0.868** | **0.894** | **0.017** |
| CVC-T | UNet | 0.710 | 0.627 | 0.684 | 0.843 | 0.876 | 0.022 |
| | UNet++ | 0.707 | 0.624 | 0.687 | 0.839 | 0.898 | 0.018 |
| | SFA | 0.467 | 0.329 | 0.341 | 0.640 | 0.817 | 0.065 |
| | PraNet | 0.871 | 0.797 | 0.843 | 0.925 | 0.972 | 0.010 |
| | **CaraNet** | **0.903** | **0.838** | **0.887** | **0.940** | **0.989** | **0.007** |

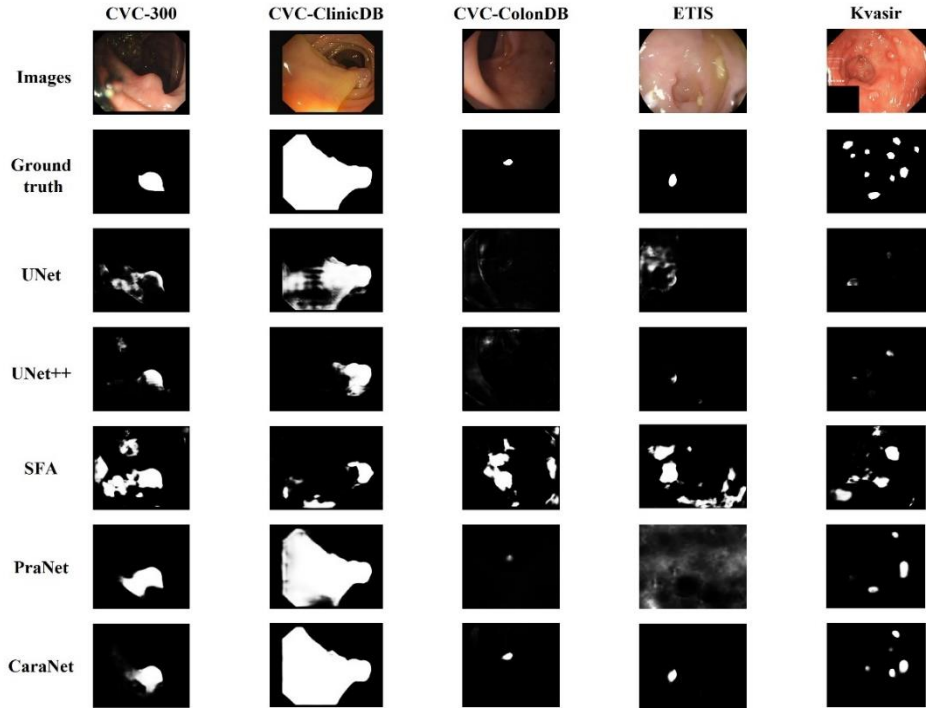

Figure 6. Polyp segmentation results

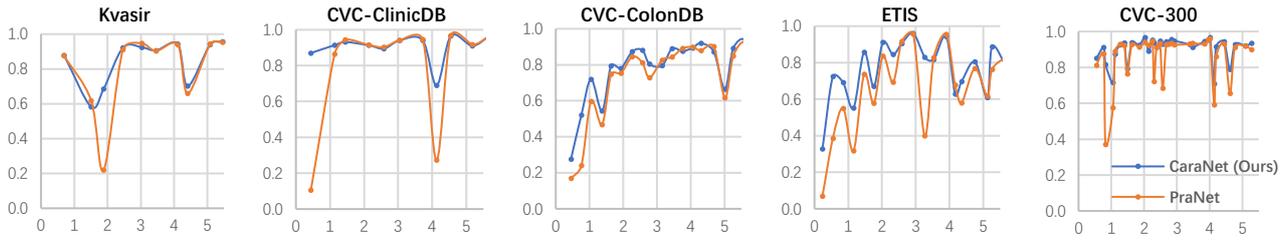

Figure 7. Performance vs. Size on the five polyp datasets. The x-axis is the proportion size (%) of polyp; y-axis is the averaged mean Dice coefficient. Blue is for our CaraNet and orange is for the PraNet.

Table 3. Quantitative results on brain tumor (BraTS 2018) dataset.

| Methods | mean Dice | mean IoU | $F_\beta^w$ | $S_\alpha$ | $E_\phi^{max}$ | MAE |
|---|---|---|---|---|---|---|
| CaraNet (Ours) | **0.631** | **0.507** | **0.629** | **0.786** | **0.927** | 0.003 |
| PraNet (MICCAI'20) | 0.619 | 0.494 | 0.606 | 0.776 | 0.920 | 0.003 |

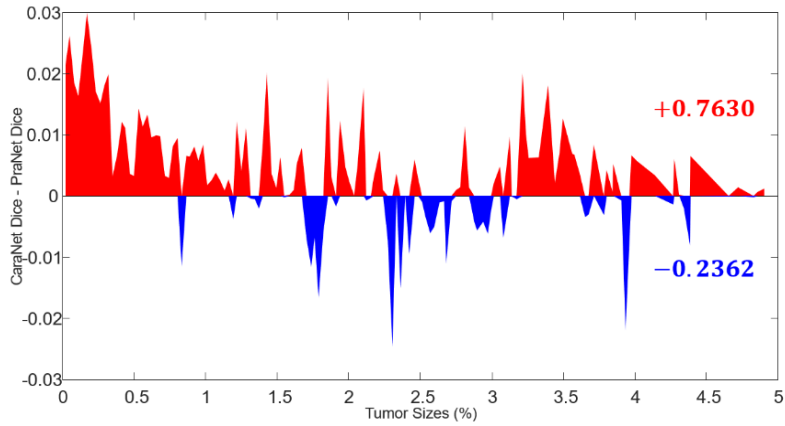

Figure 8. Performance vs. Size on brain tumor datasets. The x-axis is the proportion size (%) of tumor; the y-axis is the difference between the averaged mean Dice coefficient results of CaraNet and PraNet (CaraNet meanDice – PraNet meanDice). Red areas show CaraNet outperforms PraNet, blue areas show the opposite. The summation of red areas is 0.763 and summation of blue areas is 0.236; that means CaraNet overall outperforms PraNet.

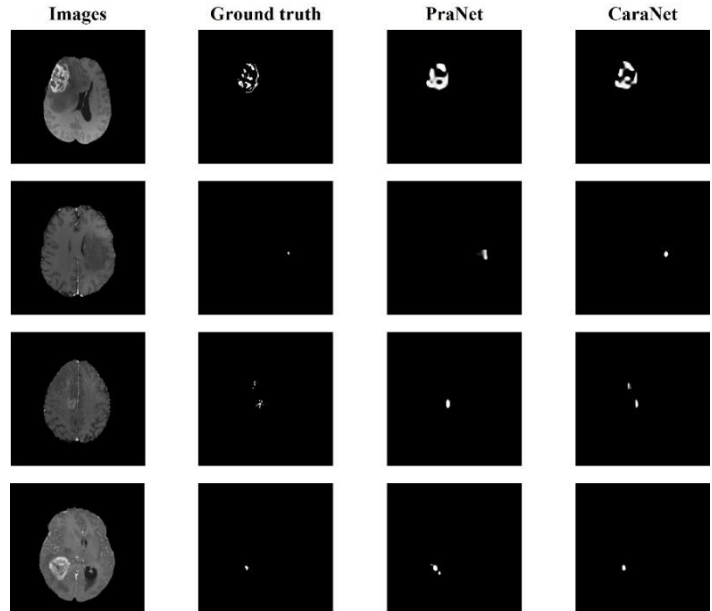

Figure 9. Brain tumor segmentation results

To further evaluate the effectiveness of CaraNet for small-object segmentation, we conducted another experiment using the brain tumor dataset (BraTS 2018). The polyp datasets lack extremely small objects (the minimum is about 0.11%) and do not have enough small samples (like Kvasir and CVC-ClinicDB, in Figure 7, there are fewer samples in the range of small sizes). The brain tumor dataset was created from the BraTS 2018 database by slicing 2D images from the "T1ce" source with "NET" type labels. We randomly select 60% of the images as the training set and the remainder as the testing dataset. Altogether, 3231 images with proportions of tumor sizes ranging from 0.01% – 4.91% were in the testing dataset. Table 3, Figure 8, and Figure 9 show the comparison result. We compared CaraNet with only PraNet for the same reason stated above. Clearly, our CaraNet performed better, especially for the extremely small cases (range 0.01% – 0.1% in Figure 8 and the red area indicates that the Dice value of CaraNet is greater than that of PraNet; blue area shows the opposite. Values on the right show the summations of all red and blue differential values).

## 4. DISCUSSION

We propose a novel deep-learning based segmentation model – CaraNet, by combining the Axial Reverse Attention and Channel-wise Feature Pyramid (CFP) modules. This new method can help improve the performance of the segmentation of small medical objects. Through the experiments, we show that CaraNet outperforms the most famous models by a large margin overall for six measurement metrics. As shown by the polyp segmentation results, CaraNet not only produces high quality segmentation on samples of large polyps, but also performs well for small and multi small-object segmentation. Figure 9 shows some results of extremely small tumor segmentation from the BraTS 2018 dataset. The advantage of the CaraNet in segmenting small single- and multi-objects is evident. In addition, compared with the recent state-of-the-art network, PraNet, CaraNet provides a more precise prediction for the most-challenging cases.

Although CaraNet achieves good improvement on medical image segmentation tasks, there are still some limitations and potentials to optimize the model. For example, using bilinear interpolation to up-sample feature maps cannot avoid the loss of some useful information and lead to a coarse boundary. It can be improved by applying a deconvolutional layer. In addition, the backbone of CaraNet is pre-trained on ImageNet, which contains natural images that are very different from medical images. Moreover, the sliced brain MRI data also cause loss of spatial information between the voxels; that may influence the accuracy of small tumor detection. In our future work, we will use the Model Genesis [47] as a 3-D backbone to replace the Res2Net (2-D backbone) and adjust the CaraNet to employ it on 3-D medical imaging segmentation to build a 3-D version CaraNet for more accurate CT or MRI image segmentation. Segmentation of 3-D medical images is of growing interest; we believe CaraNet will address that problem successfully.

We also introduce the process to evaluate segmentation models according to the size of objects. We consider the object's size using the size ratio including the sizes of objects and the whole image. In this study, we assume the size ratios of "small objects" are less than 5%. However, we did not discuss specific reasons to choose the threshold of 5%, and there is not a clear definition of the "small objects". Since few studies have fully considered the sizes of objects and the small-object problems in medical imaging, we will further study this question in future work.

## 5. CONCLUSION

We have proposed a novel neural network, CaraNet, for small medical object segmentation. From the overall segmentation accuracy, we can find that CaraNet outperforms all state-of-the-art approaches by at least 2% (mean dice accuracy). For the early diagnosis dataset (ETIS) which contains many small polyps, however, CaraNet can reach 74.7% mean dice accuracy, which is about 12% higher than PraNet. For the extremely-small object segmentation dataset (BraTS 2018), CaraNet can achieve 3% higher than PraNet. When evaluating segmentation models according to the size of objects, CaraNet outperforms PraNet for small objects in all six datasets we used.